\makeatletter
\@ifundefined{@parse@version@dash}{%
\def\@parse@version#1{\@parse@version@0#1}
\def\@parse@version@#1/#2/#3#4#5\@nil{%
\@parse@version@dash#1-#2-#3#4\@nil}
\def\@parse@version@dash#1-#2-#3#4#5\@nil{%
  \if\relax#2\relax\else#1\fi#2#3#4 }
}{}
\makeatother

\documentclass[%
reprint,
superscriptaddress,
nofootinbib,
 amsmath,amssymb,
 aps,
prl,
floatfix,
]{revtex4-2}

\usepackage{dcolumn}% Align table columns on decimal point
\usepackage{bm}% bold math

\usepackage{aas_macros}
\usepackage{xspace}
\usepackage{color}
\usepackage{hyperref}% add hypertext capabilities
\begin{document}

\preprint{APS/123-QED}

\title{A New Way to Limit the Interaction of Dark Matter with Baryons}

\author{Abraham Loeb}
\email{aloeb@cfa.harvard.edu}

\affiliation{Department of Astronomy, Harvard University, 60 Garden Street, Cambridge, MA 02138, USA}

\date{\today}

\begin{abstract}
 
Recently, there had been renewed interest in limiting the interaction
between dark matter particles and known particles. I propose a new way
to set upper limits on the coupling of ions or electrons to dark
matter particles of arbitrary mass, based on Faraday's Law in a
spinning conductor.

\end{abstract}

\maketitle

\textbf{Introduction} -- The nature of dark matter (DM) is
unknown. One way to unravel the properties of DM particles is through
their interaction with Standard Model particles, such as protons or
electrons.  Searches for weakly interacting massive particles by
direct detection experiments have set strong bounds at GeV to TeV mass
scales
\citep{PhysRevD.97.115047}.  Lighter DM particle candidates
with masses below GeV are well-motivated in a variety of particle
physics scenarios
\citep{2008PhRvL.101w1301F,2015PhRvL.115b1301H,2016PhRvL.116v1302K,2017PhRvD..96k5021K}. However,
the search for sub-GeV DM particles in traditional direct detection
experiments is limited by the suppressed momentum transfer in nuclear
recoil. Cosmological searches for light DM particles complements
direct detection experiments for DM particles down to keV masses
\citep{2002astro.ph..2496C,2014PhRvD..89b3519D,2018PhRvD..97j3530X,2018PhRvD..98h3510B,2018PhRvD..98b3013S,2021PhRvL.126i1101N,2021ApJ...907L..46M}.

The existing upper limits on the cross-section for DM particles with
protons were summarized recently in Figure 1 of
Ref. \citep{2021arXiv211110386R}, extending down to values of $\sim
10^{-30}~{\rm cm}^2$ for DM particle masses $m_{\rm dm}\sim 1$ keV.

Here, I consider a novel way to limit the proton-DM coupling based on
the magnetic field generated in a spinning conductor. For pedagogical
purposes, I illustrate the method in the context of the interstellar
medium of the Milky Way galaxy, although its most powerful
implementation should be in dedicated laboratory experiments.

\textbf{Induced Magnetic Field} -- The interstellar gas of the Milky
Way galaxy is organized into a thin disk with a flat rotation curve of
circular velocity, $v_{\rm c}\sim 240~{\rm
km~s^{-1}}$ \citep{2019ApJ...885..131R}. The ionized component of the
gas is composed primarily of electrons and
protons\citep{2011piim.book.....D,1998ppim.book.....S}. The Galactic
disk is immersed in a DM halo with a characteristic mass density,
$\rho_{\rm dm}\sim 0.5~{\rm GeV~cm^{-3}}$, in the vicinity of the
Sun \citep{2020MNRAS.495.4828G}. 

For a proton-DM cross-section $\sigma$, the proton-DM collision
frequency is: $\nu_{\rm p-dm} \sim (\rho_{\rm dm}/m_{\rm
dm})\mu \sigma v_{\rm c}$, where the scattering kinematics implies
$\mu \sim (m_{\rm dm}/m_{\rm p})$ for $m_{\rm dm}\ll m_{\rm p}$ and
$\mu \sim 1$ for $m_{\rm dm}\gg m_{\rm p}$.  The proton-DM coupling could
also be associated with a long-range interaction that is not mediated by
two-particle collisions. We therefore derive a general limit on
$\nu_{\rm p-dm}$, independent of the nature of the interaction.

The momentum change of protons as a result of scattering on DM
particles over a timescale $1/\nu_{\rm p-dm}$ results in a drag force
per unit volume on the plasma: $-(n_{\rm p}m_{\rm p}\nu_{\rm
p-dm}){\vec v}_{\rm c}$, where $m_{\rm p}$ and $n_{\rm p}$ are the
proton mass and number density. This force generates a longitudinal electric
field ${\vec E}$ along the direction of motion, which carries the
electrons and protons together at the same bulk velocity ${\vec{v_{\rm
c}}}$ and maintains local quasi-neutrality of the
plasma \citep{1988PhRvD..37.3484L},
\begin{equation}
e{\vec E} = m_{\rm e}\nu_{\rm p-dm}{\vec v}_{\rm c} ,
\label{eq:1}
\end{equation}
where $m_e$ is the electron mass. 

For a flat rotation curve along the radial coordinate $r$ in
cylindrical symmetry, we get ${\vec \nabla}\times {\vec v}_{\rm
c}=(1/r)[\partial(rv_{\rm c})/\partial r]= (v_{\rm c}/r) {\hat e}_z$,
in the $z$-direction perpendicular to the disk plane. As in the
Biermann Battery case \citep{1950ZNatA...5...65B}, Faraday's Law implies
the generation of a vertical magnetic field ${\vec B}=B{\hat e}_z$ at
a rate,
\begin{equation}
{\partial {\vec B}\over \partial t}= - c {\vec \nabla}\times {\vec E}=
-\left({c m_{\rm e}\over e}\right)\nu_{\rm p-dm}
\left({v_{\rm c} \over r}\right) {\hat e}_z .
\label{eq:2}
\end{equation}

Coulomb collisions do not erase a large-scale electric field. They
only affect the current driven in the plasma through the relation
between electric field, ${\vec E}$, and conductivity, $\sigma_c$ (which
is proportional to the electron-proton collision time) through Ohm's
law: ${\vec j}=\sigma_c({\vec E}+{\vec v}\times {\vec B})$. The
resulting limit on the electric current does not limit the build-up of
the magnetic field. In fact, if Faraday's Law was not used to impose a
limit on the magnetic field under the quasi-neutrality of the plasma,
then the characteristic current, ${\vec j}$, driven by the frictional
force on the dark matter would have resulted in a much larger magnetic
field based on the other Maxwell equation: ${\vec \nabla}\times {\vec
B} - (d{\vec E}/dt)=(4\pi/c){\vec j}$.

Over the age of the Milky Way disk, $t_{\rm d}\sim 10^{10}$yr, the
magnetic field builds up to a magnitude,
\begin{equation}
B= \left({c m_{\rm e}\over e}\right) \left({v_{\rm c} \over
r}\right)\nu_{\rm p-dm} t_{\rm d} .
\label{eq:3}
\end{equation}
The net effect on the plasma originates because of its net bulk motion
relative to the dark matter. Collisions occur at an equal rate from
all directions if the velocity distribution of the dark matter
particles is isotropic and the plasma has no bulk motion relative to
the dark matter. But the Milky-Way disk has a net rotation relative to
the dark matter halo in which it is embedded, generating the
above-mentioned magnetic field. The collisions between protons and
electrons in the plasma are not associated with a bulk motion of the proton
fluid relative to the electron fluid. In fact, quasi-neutrality
implies that they move together hydrodynamically. This is achieved
through the large-scale electric field in equation (\ref{eq:1}).

Substituting all the above expressions for the fiducial parameters of
the Milky Way disk and halo at the Galactocentric radius of the Sun,
$r\sim 8~{\rm kpc}$, yields a magnetic field vertical to the disk plane
of magnitude, 
\begin{equation}
B\sim 2 \mu G \left({\nu_{\rm p-dm}\over 0.1~{\rm s}^{-1}}\right).
\label{eq:4}
\end{equation}

The measured value of the interstellar magnetic field of a few
$\mu$G \citep{2011piim.book.....D,2017A&A...603A..15O} places an upper
limit on the proton-DM interaction for arbitrary DM particle
masses, namely  ${\nu_{\rm p-dm}\lesssim 0.1~{\rm s}^{-1}}$

Laboratory experiments in superconductors or plasmas on sub-micron
scales with $r\lesssim 10^{-4}$cm, $v_c\sim c$, $t_{\rm d}\gtrsim 1$yr
and sensitivity to $B\sim 1$nG, could potentially constrain
unprecedented levels of $\nu_{\rm p-dm}\lesssim 10^{-24}~{\rm
s}^{-1}$, corresponding to $\sigma \lesssim 10^{-34}~{\rm cm}^2$ for
arbitrarily low DM particle masses, $m_{\rm DM}\ll 1~{\rm GeV}$.  Such
limits would require exquisite control over noise from interactions
with other particles or waves in the environment of the
experiment. The experimental design of an optimal laboratory set-up
based on this novel detection method should be explored in the future.

\textbf{Conclusions} -- Equations (\ref{eq:1}-\ref{eq:3}) can be used to
limit $\nu_{\rm p-dm}$ for arbitrary DM particle masses and an
observed magnetic field in a spinning conductor.  The limit also
applies to any long-range force between DM and protons.

In deriving this limit we focused on the interaction of protons with
DM particles. An analogous limit can be derived by considering any
preferential interaction of electrons with DM particles. It is
unlikely that the DM particles would produce an identical frictional
force density for electrons and ions in a moving conducting medium
because of the different masses and couplings of leptons and quarks.

\textbf{Acknowledgments} -- This work was supported in part by a grant
from the Breakthrough Prize Foundation.

\bibliography{pDM}% Produces the bibliography via BibTeX.
\end{document}